\documentclass[authoryear,preprint,review,12pt]{elsarticle}
\usepackage[flushleft]{threeparttable}
\usepackage{graphicx}

\usepackage{amssymb}
\usepackage{amsthm}

\usepackage{lineno}

\usepackage{longtable}
\usepackage{colortbl}

\journal{Icarus}

\begin{document}

\hyphenpenalty=10000\relax 
\exhyphenpenalty=10000\relax 
\sloppy

\begin{frontmatter}



\title{Photometric Observations of 107P/Wilson-Harrington}


\author[label1]{Seitaro~Urakawa}
\author[label1]{Shin-ichiro~Okumura}
\author[label1]{Kota~Nishiyama}
\author[label1]{Tsuyoshi~Sakamoto}
\author[label1]{Noritsugu~Takahashi}
\author[label2]{Shinsuke~Abe}
\author[label3]{Masateru~Ishiguro}
\author[label4]{Kohei~Kitazato}
\author[label5]{Daisuke~Kuroda}
\author[label6]{Sunao~Hasegawa}
\author[label7]{Kouji~Ohta}
\author[label8]{Nobuyuki~Kawai}
\author[label5]{Yasuhiro~Shimizu}
\author[label9]{Shogo~Nagayama}
\author[label5]{Kenshi~Yanagisawa}
\author[label10]{Michitoshi~Yoshida}
\author[label1,label6]{Makoto~Yoshikawa}

\address[label1]{Bisei Spaceguard Center, Japan Spaceguard Association, 1716-3 Okura, Bisei, Ibara, Okayama 714-1411, Japan}
\address[label2]{Institute of Astronomy, National Central University, 300 Jhongda Road, Jhongli, Taoyuan, 32001, Taiwan}
\address[label3]{Department of Physics and Astronomy, Seoul National University, 599 Gwanak-ro, Gwanak-gu, Seoul 151-742, Republic of Korea}
\address[label4]{Research Center for Advanced Information Science and Technology, University of Aizu, Aizu-Wakamatsu, Fukushima 965-8580, Japan}
\address[label5]{Okayama Astrophysical Observatory, NAOJ, 3037-5 Honjo, Kamogata, Asakuchi, Okayama 719-0232, Japan}
\address[label6]{Institute of Space and Astronautical Science, Japan Aerospace Exploration Agency, 3-1-1 Yoshinodai, Chuo-ku, Sagamihara, Kanagawa 252-5210, Japan}
\address[label7]{Department of Astronomy, Kyoto University, Sakyo-ku, Kyoto 606-8502, Japan}
\address[label8]{Department of Physics, Tokyo Institute of Technology, 2-12-1 Ookayama, Meguro-ku, Tokyo 152-8551, Japan}
\address[label9]{National Astronomical Observatory of Japan, 2-21-1 Osawa, Mitaka, Tokyo 181-8588, Japan}
\address[label10]{Department of Physical Science, Hiroshima University, 1-3-1 Kagamiyama, Higashi-Hiroshima, Hiroshima 739-8526, Japan}

\begin{abstract}

We present lightcurve observations and multiband photometry for 107P/Wilson-Harrington using five small- and medium-sized telescopes. The lightcurve has shown a periodicity of $0.2979$ $day$ (7.15 $hour$) and $0.0993$ $day$ (2.38 $hour$), which has a commensurability of 3:1. The physical properties of the lightcurve indicate two models: (1) 107P/Wilson-Harrington is a tumbling object with a sidereal rotation period of $0.2979$ $day$ and a precession period of $0.0993$ $day$. The shape has a long axis mode (LAM) of $L_{1}$:$L_{2}$:$L_{3}$ = 1.0:1.0:1.6. The direction of the total rotational angular momentum is around $\lambda=310^{\circ}$, $\beta=-10^{\circ}$, or $\lambda=132^{\circ}$, $\beta=-17^{\circ}$. The nutation angle is approximately constant at $65^{\circ}$. (2) 107P/Wilson-Harrington is not a tumbler. The sidereal rotation period is $0.2979$ $day$. The shape is nearly spherical but slightly hexagonal with a short axis mode (SAM) of $L_{1}$:$L_{2}$:$L_{3}$ = 1.5:1.5:1.0. The pole orientation is around $\lambda=330^{\circ}$, $\beta=-27^{\circ}$. In addition, the model includes the possibility of binary hosting. For both models, the sense of rotation is retrograde. Furthermore, multiband photometry indicates that the taxonomy class of 107P/Wilson-Harrington is C-type. No clear rotational color variations are confirmed on the surface. 

\end{abstract}

\begin{keyword}
Asteroids rotation \sep Comets \sep Photometry \sep Satellites of asteroids

\end{keyword}

\end{frontmatter}
\renewcommand{\figurename}{Fig.}
\renewcommand{\tablename}{Table}



\section{Introduction}
Asteroids and comets are primordial bodies that formed in the earliest stage of the solar system. Their rotational states, shapes, and material reflect the collisions, disruptions, and chemical processes since then to the present. Some small solar system bodies exhibit behavior such as that shown by both comets and asteroids (so-called, comet-asteroid transition objects). As an example, near-earth object (NEO) (3200) Phaethon shows signs of past cometary activity because it is thought to be associated with the Geminid \citep{Gustafson89}. Dynamical numerical simulations and spectral observations for (3200) Phaethon support (2) Pallas, which is outer main belt asteroids, is the most likely parent body of (3200) Phaethon (\citealt{Clark10}; \citealp{deLeon10}). Meanwhile, objects that display cometary activities in the main-belt asteroid (MBA) region have recently been discovered. They are classified as  main-belt comets (MBCs) \citep{Hsieh06a}; the MBCs are 133P/Elst-Pizzaro \citep{Elst96}, P/2005 U1 \citep{Read05}, 176P/LINEAR \citep{Hsieh11}, P/2008 R1 (Garrad) \citep{Jewitt09},  P/2010 A2 (LINEAR) \citep{Birt10}, P/2010 R2 (La Sagra) \citep{Marsden10}, and (596) Scheila (\citealp{Bodewits11}; \citealp{Jewitt11}). One possible activation mechanism for MBCs is impacts with small (e.g., meter-sized) objects (\citealp{Toth00}; \citealp{Diaz08}; \citealp{Jewitt10}; \citealp{Snodgrass10}; \citealp{Bodewits11}; \citealp{Jewitt11}). The other activation mechanisms are  rotational-fissions due to the spin-up by Yarkovsky-O'Keefe-Radzievskii-Paddack (YORP) effects \citep{Jewitt10}, and thermal influences \citep{Jewitt09}. Interesting properties of  MBCs are their dynamical origin and possible function as reservoirs for water-ice and organics. A numerical integration by \cite{Haghighipour09} states that the origin of 133P/Elst-Pizarro, 176P/LINEAR, and P/2005 U1 (Read) is concordant with the Themis family of asteroids.  Compared with all asteroids, the Themis family of asteroids includes B-type asteroids at a relatively high population rate. Some B-type asteroids in the Themis family seem to have experienced aqueous alterations (\citealp{Yang10}; \citealp{Clark10}). (3200) Phaethon is also a B-type asteroid and shows the existence of aqueous alteration materials \citep{Licandro07}. In addition, water-ice and organics are detected on the surface of (24) Themis \citep{Rivkin10} and (65) Cybele, which orbits along the outer edge of the main belt \citep{Licandro11}. The study of comet-asteroid transition objects provides keys to the dynamical origin and evolution of NEOs, the mutual collisions of small solar system bodies, the material differences between asteroids and comets, and the origin of Earth's water. 

 This study's purpose is to obtain the rotational states, shape model, and rotational color variations for 107P/Wilson-Harrington (also know as (4015) Wilson-Harrington; hereafter 107P), which is a representative comet-asteroid transition object. 107P was discovered accompanied by a faint cometary tail at Palomar Observatory in 1949 \citep{Fernandez97}. The object, however, could not be tracked because of insufficient observations to determine an accurate orbit. Later, a near-earth asteroid 1979VA (= 4015) was discovered. Subsequent observations identified asteroid (4015) 1979VA and 107P as the same object. Despite a devoted search, no cometary activity has been detected since the initial observation of 107P (\citealp{Chamberlin96}; \citealp{Lowry03}; \citealp{Ishiguro11}). 107P is an Apollo asteroid whose orbital parameters are a = 2.639 AU, e = 0.624, i = 2.785$^{\circ}$, and the Tisserand
parameters ($T_{J}$) = 3.08. A numerical simulation by \citet{Bottke02} mentions that  there is a 4 \% chance that 107P has a JFC origin and a 65\% chance it has an origin in the outer main-belt region. Taxonomically, it is categorized as a CF-type \citep{Tholen89}.  The reflectance spectrum in the region 3800--6200 {\AA} is similar to (3200) Phaethon \citep{Chamberlin96}.  The thermal properties of 107P have been investigated by mid-infrared photometry with NASA's Spitzer Telescope \citep{Licandro09}. These observations show that the beaming parameter, the diameter, and the albedo are $\eta$ = 1.39 $\pm$ 0.26, $D$ = 3.46 $\pm$ 0.32 $km$, and $p_{v}$ = 0.059 $\pm$ 0.011, respectively. 
   The rotational period of 107P has been reported to be 3.556 $hour$ and 6.10 $\pm$ 0.05 $hour$ by \citet{Harris83} and \citet{Osip95}, respectively. \citet{Osip95} ascribes the difference of the two reports to the noisy data of \citet{Harris83} because of the weather conditions.  The few days' observation in both reports, however, is not enough to determine the correct rotational period. Longer observations are required to derive the correct rotational period and other physical properties.
 
We hypothesize that 107P migrates to the NEO region from the outer main-belt region inhabited by six of seven known MBCs, and impacts with small objects could eject dust and/or expose sub-surface ice that then trigger 107P's cometary activity. Post-MBC, 107P is capable of becoming host to water-ice, organics, and aqueous alteration materials. In this hypothesis, the impacts' influence would be apparent in the rotational states and/or the surface color variations.

We had an opportunity to observe 107P from August 2009 to March 2010. Our long observation campaigns enable us to derive the rotational states, shape model, and rotational color variations. Furthermore, the orbit of 107P makes it accessible by spacecraft.  A more advanced sample return mission from a D-type asteroid or an asteroid-comet transition object is envisioned in Japan. One candidate is 107P \citep{Yoshikawa08}. Clarification of the physical properties of 107P is important to the design of the future mission. If we are able to obtain 107P's physical properties, the data will be useful to revise the physical model of \cite{Licandro09}, similar to the Hayabusa-2 target 162173 (1999JU3)  whose physical model was reconstructed by both thermal observations and the lightcurve \citep{Muller11}. This paper is organized as follows. In Section 2, we describe the observations made and the data reduction. In Section 3, we mention the rotational states and shape model of 107P. In Section 4, we focus on the possibility of tumbling motion and the existence of a binary. Finally, we summarize the physical model of 107P and discuss the feasibility of a sample return mission. 

\section{Observations and Data Reduction}
\subsection{Observations}
We conducted the photometric observation campaigns of 107P with five small- and medium-sized telescopes. The observational circumstances and the states of 107P are listed in Table 1 and Table 2, respectively. All telescopes were operated with the non-sidereal tracking mode. The longest-term observation of this campaign was carried out using the 1.0 m f/3 telescope at the Bisei Spaceguard Center (BSGC\footnote{BSGC is administrated by the Japan Space Forum.}) from September 6, 2009 to March 11, 2010. The detector consisted of four CCD chips with 4096 $\times$ 2048 pixels. We used one CCD chip to obtain as many images as possible by shortening the processing time. The field of view (FOV) for one CCD chip is $1.14{^\circ}$  $\times$ $0.57{^\circ}$ with a pixel resolution of 1.0$''$. The exposure time varied from 30 $s$ to 600 $s$ according to the observational situations. Individual images were taken with a commercially available short-pass (long-wavecut) filter, the effective wavelength of which ranged from 490 $nm$ to 910 $nm$. We denote the filter as $W$ in Table 1. In order to investigate rotational color variations, multiband photometry was conducted using a Sloan Digital Sky Survey (SDSS) $g'$, $r'$, $i'$, $z'$ filter on December 17, 2009. One set of observations was made using three consecutive images for each filter. The filters were changed in the following sequence: three $g'$ images ${\rightarrow}$ three $r'$ images ${\rightarrow}$ three $i'$ images ${\rightarrow}$ three $z'$ images. We repeated this sequence four times.

The second-longest-term observation used the 0.5 m f/6.5 Multicolor Imaging Telescope for Survey and Monstrous Explosions (MITSuME) \citep{Kotani05} at Okayama Astrophysical Observatory (OAO) from November 7, 2009 to December 21, 2009. The telescope is capable of obtaining a three-color (SDSS $g'$, Johnson--Cousins $R_{c}$ and $I_{c}$) image simultaneously. The detector is 1024 $\times$ 1024 pixels CCD with  FOV of  $26'$ $\times$ $26'$ (1.52$''$/pixel). The images were taken with an exposure time of 120 $s$. To search for the rotational color variation, we used the data of December 17 because they could be compared with the observations of the BSGC and the photometric precision of the other day's data was not sufficient to detect the color variation. 

The third observation was carried out using a 1.05 m f/3.1 Schmidt telescope with 2048 $\times$ 2048 pixels CCD at Kiso Observatory on August 17, 19, and 20 and December 12, 2009. This instrument provides a FOV of $50' $ $\times$ $50'$ (1.46$''$/pixel). The images were obtained using a  Kron--Cousins $R_{c}$ filter with an exposure time of 120--300 $s$. 

The fourth observation was made using the Lulin One-meter  Telescope (LOT)  \citep{Huang05} in Taiwan on December 7--10, 2009. The CCD consists of a 1340 $\times$ 1300 array, and the FOV covers the area of $11.5'$ $\times$ $11.2'$. The pixel resolution and f-number are 0.51$''$/pixel and 8, respectively. The images were obtained using a Johnson--Cousins $R_{c}$ filter with an exposure time of 90 $s$. 

The last, observation was made using the University of Hawaii 2.24 m f/10 telescope (UH88) on December 19, 2009, with a 2048 $\times$ 2048 pixels CCD. The FOV of the instrument is $7.5'$ $\times$ $7.5'$ with a pixel resolution of 0.44$''$.  Almost all images were obtained using a Kron--Cousins  $R_{c}$ filter with an exposure time of 60 $s$.

\begin{scriptsize}
\begin{threeparttable}
\begin{longtable}[c]{cccc} 
\caption{Observation states.}
\\
\hline
Observatory & Year/Mon/Day& Exp time(s) &  Filter  \\ 
\hline 
\endhead
\hline
\endfoot
\hline
\endlastfoot
\hline
BSGC & 2009/12/17 &300 &$g'$, $r'$, $i'$, $z'$ \\
BSGC\tnote{${\dagger}$} & 2009/09/6,7,9,10,15,16,19, 2009/10/8,10,28   & 30--180 & $W$ \\
BSGC & 2009/11/3,5--7,11,14, 2009/12/5,7--9,19,22 & 60--300& $W$  \\
BSGC & 2010/01/3,6--8,14--18,22,23 & 180--300&$W$\\
BSGC\tnote{${\dagger}$} & 2010/02/3--5,7,9,16,18,19, 2010/03/11 & 240--600&$W$\\
OAO & 2009/11/7,14,15,18--21,23,   & 120 & $g'$, $R_{c}$, $I_{c}$  \\
OAO & 2009/12/1,2,6,7,14,16--21   & 120& $g'$, $R_{c}$, $I_{c}$\\
KISO\tnote{${\dagger}$} & 2009/08/17,19,20, 2009/12/12    & 120--300 & $R_{c}$  \\
LOT & 2009/12/7--10     & 90 & $R_{c}$   \\ 
UH88 & 2009/12/19         & 60 & $R_{c}$   \\ 
\hline
\end{longtable}
\begin{tablenotes}
\item[${\dagger}$]A sufficient number of data was not obtained from August to October because the altitude of 107P fell below 25$^{\circ}$ about 30 minutes from the observation start and 107P overlapped stars of the galactic plane. The photometric precision was insufficient after February 7, 2010. We did not use these data for the estimation of rotational periods and shape models.  These data were utilized for the trend  confirmation of lightcurves that were obtained from the other day's data and the monitoring of cometary activity. 
\end{tablenotes}
\end{threeparttable}
\end{scriptsize}

\begin{scriptsize}
\begin{threeparttable}
\begin{longtable}[c]{cccccc} 
\caption{States of 107P in each month.}
\\
\hline
Year/Mon/Day  &  $\Delta$ [AU]\tnote{a}& R [AU]\tnote{b}  & $\alpha$ [$^{\circ}$]\tnote{c} & Sky motion [$''$/min] & $m$ \tnote{d}  \\ 
\hline 

\endhead
\hline
\endfoot
\hline
\endlastfoot
\hline
2009/08/17--20 & 0.687--0.684& 1.309--1.286& 49.8--51.4& 0.45--0.55 & 17.7--17.7\\
2009/09/6--19   & 0.653--0.612& 1.162--1.084& 59.9--66.0&1.09--1.49 &17.9--17.9\\
2009/10/8--28 & 0.529--0.434& 1.008--0.995& 73.8--77.2&2.18--3.26 &17.3--17.5\\
2009/11/3--23 & 0.410--0.382&1.006--1.083 &76.3--65.4& 3.65--4.52 &17.1--16.7\\
2009/12/1--22 & 0.401--0.543& 1.130--1.277& 59.2--46.3&4.31--2.96 &16.6--17.0\\
2010/01/3-23  & 0.670--0.936& 1.373--1.539& 42.0--37.8& 2.32--1.70 & 17.4--18.4\\
2010/02/3--19 & 1.105--1.372&1.632--1.767 & 36.1--33.8&1.51--1.34 &18.8--19.7\\
2010/03/11     & 1.732& 1.932& 30.8&1.22&20.0\\
\hline
\end{longtable}
\begin{tablenotes}
\item[a]Object to observer distance.
\item[b]Heliocentric distance.
\item[c]Phase angle (Sun--107P--observer).
\item[d]Apparent magnitude. This value is estimated using UCAC 2 catalog stars that are taken in the same field with 107P.
\end{tablenotes}

\end{threeparttable}
\end{scriptsize}

 \subsection{Data reduction}
All images were bias and flat-field corrected. When using the data of OAO for the derivation of lightcurve, we stacked four images to compensate for the poor flux. All observation times were corrected using the light travel time from 107P to the Earth. By using the IRAF/APPHOT\footnote{IRAF is distributed by the National Optical Astronomy Observatory, which is operated by the Association of Universities for Research in Astronomy (AURA) under cooperative agreement with the National Science Foundation.} package, we measured the raw magnitude of 107P and from three to seven reference stars that were bright enough compared with 107P. We set aperture radius to 1.5 ${\times}$ FWHM for 107P and reference stars images, respectively. Since the reference star images are slightly elongated by the non-sidereal tracking, the aperture radius  is larger than that of 107P.  We calibrated the magnitude fluctuations due to the change of atmospheric conditions as follows,

\begin{equation}
F_{c}^{i}(t)=F_{o}^{i}(t)-\overline{F_{r}^{i}(t)}.
\end{equation}
Here, $F_{c}^{i}(t)$ is the lightcurve by rotation of 107P in $i$-th observation day, $F_{o}^{i}(t)$ is the raw magnitude of 107P, $\overline{F_{r}^{i}(t)}$ is the averaged raw magnitude of reference stars and represents the change of atmospheric conditions, and $t$ is the observational time. Next,  we define the averaged magnitude of $F_{c}^{i}(t)$ in each night as the normalized (zero) magnitude. The lightcurve by rotation of 107P can be rewritten as

\begin{equation}
F_{wh}^{i}(t)=F_{c}^{i}(t)-\overline{F_{c}^{i}},
\end{equation}
where $\overline{F_{c}^{i}}$ is the averaged magnitude of $F_{c}^{i}(t)$. Since the averaged magnitude is normalized to zero magnitude for all nights, we can connect  the different night's lightcurve with little regard for the difference of absolute magnitude. In addition, the difference of reference stars each night does not affect the periodicity of lightcurve. The problem of this procedure could include the offset between different nights when the short observation time per day poses the detection of a specific peak (bottom) in the lightcurve. However, the peaks and bottoms in the lightcurve have been detected evenly (See Fig. 3) because the observation time per day is long enough from November 2009 to February 2010 when the data are utilized for the analysis. Thus, the offset is negligible. Furthermore, the apparent magnitude change of 107P is gradual up to 1.0 magnitude per month (Table 2). The change does not act on the derivation of rotational period that is expected to be from three to seven hours according to past reports.

In contrast to the relative photometry of the lightcurve, more photometric precision is required to detect the rotational color variation by multiband photometry. In order to improve the photometric precision, we averaged three consecutive images of 107P for the BSGC's data and 14--16 consecutive images of 107P for the OAO's data. We also measured the flux of ten standard stars from SDSS data Release 7 \citep{Abazajian09}, whose stars were imaged simultaneously in the same frame as 107P (Table 3). These objects have magnitudes of about 14 $mag$ to 16 $mag$ in the $r'$-band and classification code 1 (= primary), quality flag 3 (= good), and object class 6 (= star). We evaluated atmospheric extinction coefficients and conversion factors to standardize the SDSS system for each filter. The atmospheric extinction coefficient was calculated by the magnitude variations of the standard stars for the change in airmass. Extra-atmospheric instrumental magnitudes of both 107P and the standard stars were derived using the obtained atmospheric extinction coefficient. The conversion factors were estimated by comparing the extra-atmospheric instrumental magnitudes with the magnitude of standard stars. In BSGC's observation, the multi-color images were not obtained simultaneously. The brightness of 107P by the rotation changes inevitably during the filter switch. We defined the time of the third $r' $ images in each sequence as a standard time, and then calibrated the amount of brightness change for the standard time. The amount of brightness change was estimated by the fitting curve of the lightcurve. In order to compare the OAO's data with BSGC's, the $R_{c}$ and $I_{c}$ magnitudes obtained at OAO were converted to $r'$ and $i'$ magnitudes using the conversion equations proposed by \cite{Jordi06}. 

\begin{scriptsize}
\begin{longtable}[c]{cccccc} 
\caption{Standard stars in SDSS-7.}
\\
\hline
 Ra[${^\circ}$] &Dec[${^\circ}$]  & $g'$ & $r'$ &  $i'$  &  $z'$   \\ 
\hline 
12.051081 & 8.615352 & 15.891 ${\pm}$ 0.003 & 15.279 ${\pm}$ 0.003  & 15.057 ${\pm}$ 0.003 &14.913 ${\pm}$ 0.005     \\ 
12.246608 & 8.604861 & 15.209 ${\pm}$ 0.003 & 14.756 ${\pm}$ 0.003  & 14.598 ${\pm}$ 0.003 &14.525 ${\pm}$ 0.004     \\ 
12.073952 & 8.515561 & 14.536 ${\pm}$ 0.003 & 14.215 ${\pm}$ 0.003  & 14.115 ${\pm}$ 0.003 &14.074 ${\pm}$ 0.004     \\ 
12.011267 & 8.489718 & 15.766 ${\pm}$ 0.004 & 14.891 ${\pm}$ 0.004  & 14.587 ${\pm}$ 0.003 &14.444 ${\pm}$ 0.004     \\ 
12.275938 & 8.545386 & 16.158 ${\pm}$ 0.003 & 15.566 ${\pm}$ 0.003  & 15.349 ${\pm}$ 0.004 &15.230 ${\pm}$ 0.005     \\ 
11.898784 & 8.539119 & 14.947 ${\pm}$ 0.003 & 14.554 ${\pm}$ 0.003  & 14.418 ${\pm}$ 0.003 &14.366 ${\pm}$ 0.004     \\ 
12.365509 & 8.569521 & 14.587 ${\pm}$ 0.003 & 14.037 ${\pm}$ 0.003  & 13.839 ${\pm}$ 0.003 &13.713 ${\pm}$ 0.003     \\ 
12.374946 & 8.538940 & 16.005 ${\pm}$ 0.003 & 15.555 ${\pm}$ 0.003  & 15.396 ${\pm}$ 0.004 &15.307 ${\pm}$ 0.005     \\ 
11.796300 & 8.569376 & 15.308 ${\pm}$ 0.003 & 14.641 ${\pm}$ 0.003  & 14.413 ${\pm}$ 0.003 &14.287 ${\pm}$ 0.004     \\ 
12.371825 & 8.478789 & 15.763 ${\pm}$ 0.004 & 15.298 ${\pm}$ 0.004  & 15.156 ${\pm}$ 0.004 &15.099 ${\pm}$ 0.005     \\ 
\hline
\end{longtable}
\end{scriptsize}

\section{Results}
\subsection{Rotational states}
Since the ``Standard Feature ($SF$)'' that is, the flux peaks and/or bottoms in lightcurves, shifts along the phase of lightcurves due to changes in the geometric relationship between the Earth, 107P, and the Sun during the long-term observations, the estimation of the sidereal rotation period for 107P requires a short-term observation within a few weeks. We use the data from December 7 to 22, when the phase-shift is small. The photometric precision of 107P and the observational implementation time per day are enough to make clear the sidereal rotational period during the term. Assuming double-peak lightcurves, a period analysis is carried out with a Lomb--Scargle periodgram (\citealp{Lomb76}; \citealp{Scargle82}). The power spectrum from the period analysis shows four period candidates of $0.0993$ $day$, $0.2294$ $day$, $0.2591$ $day$ and $0.2979$ $day$ (Fig. 1). 
\begin{figure}
\begin{center}
\begin{tabular}{c}
{\includegraphics[width=8cm,clip]{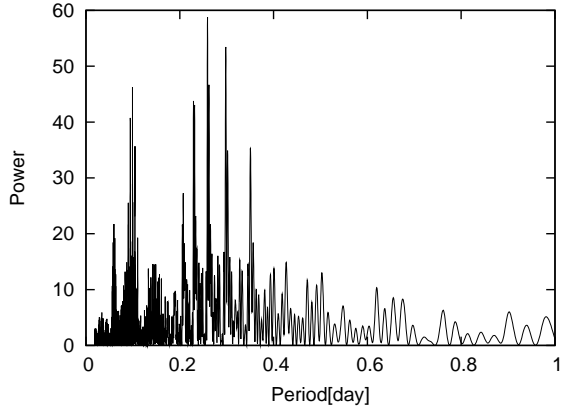}}
\end{tabular}
\end{center}
\caption{Power spectrum for the sidereal rotation period of 107P by assuming the double-peak lightcurve. The calculation is carried out on data obtained from December 7 to 22.}
\end{figure}
A typical error of  $0.0002$ $day$ corresponds to ${\pm}$ $0.005$ $hour$. Though the most significant candidate is $0.2591$ $day$, we conclude that $0.2979$ $day$  (${\simeq}$ $7.15$ $hour$) is the sidereal rotation period of 107P for the following reasons. First, the amount of the amplitude in the folded lightcurve with $0.2591$ $day$ is not stable in the same phase. That is to say, the different amplitudes overlap on a specific phase. For example, the lightcurve peaks and bottoms overlap around the phase of 0.2--0.4 in the folded lightcurve with 0.2591 $day$ (Fig. 2: Top). In the case of the folded lightcurve with $0.2979$ $day$, the same amplitudes appear periodically (Fig. 2: Bottom. A few lightcurves each night are also shown in Fig. 3). 
\begin{figure}
\begin{center}
\begin{tabular}{c}
{\includegraphics[width=8cm,clip]{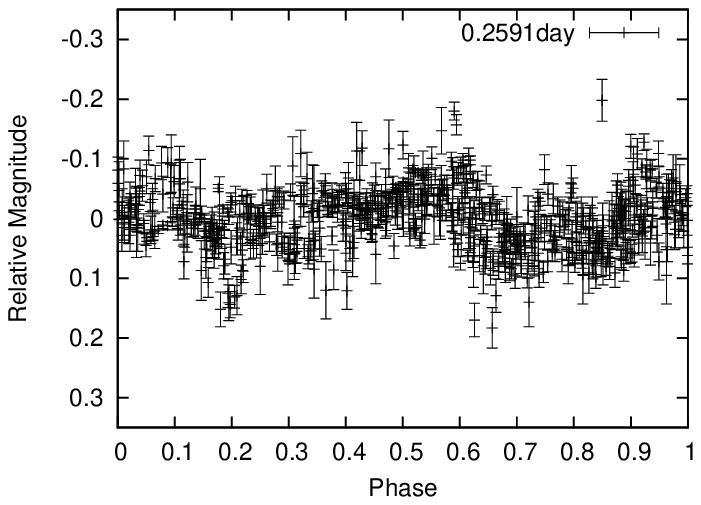}}\\
{\includegraphics[width=8cm,clip]{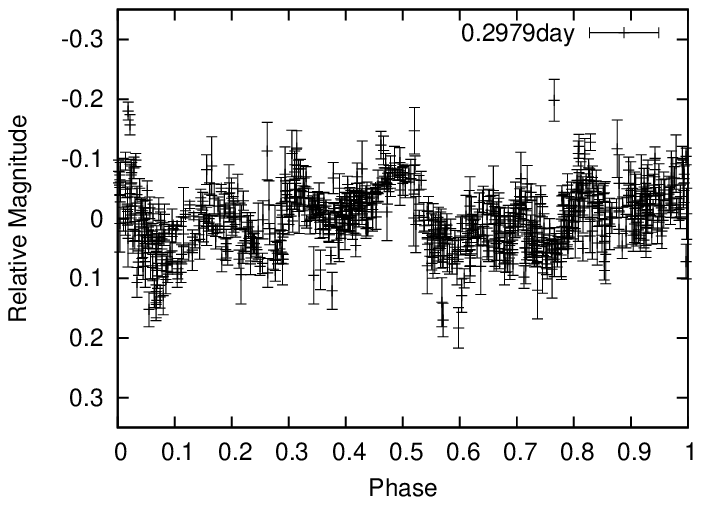}}\\


\end{tabular}
\end{center}
\caption{Lightcurve of 107P. (Top) The lightcurve is folded with 0.2591 $day$. The peak and bottom of the lightcuve overlap around the phase between 0.2 and 0.4. (Bottom) The lightcurve is folded with 0.2979 $day$. The same amplitudes appear periodically.}
\end{figure}
\begin{figure}
\begin{center}
\begin{tabular}{c}
{\includegraphics[width=8.3cm,clip]{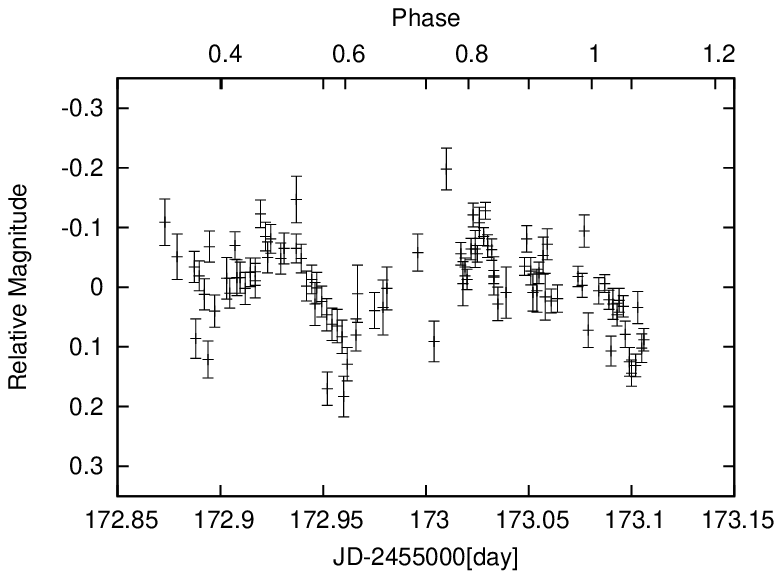}}\\
{\includegraphics[width=8.3cm,clip]{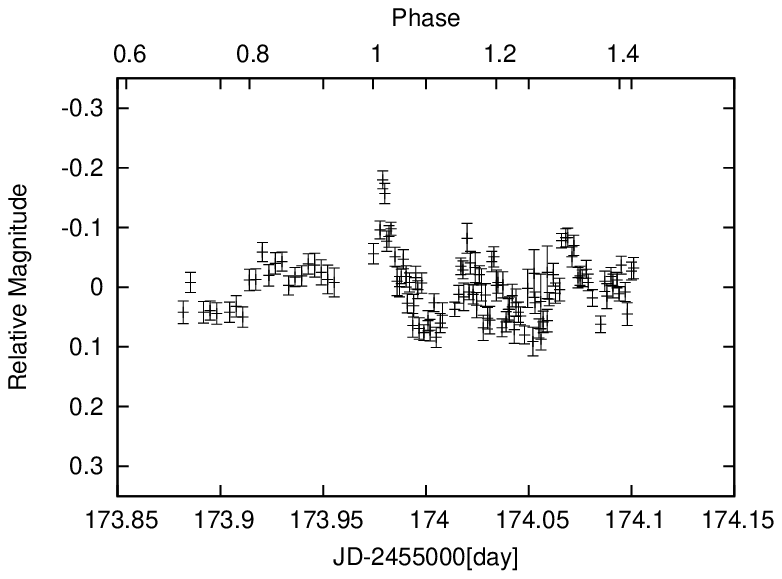}}\\


\end{tabular}
\end{center}
\caption{(Top) Lightcurve in December 7, 2009. (Bottom) Lightcurve in December 8, 2009. The phase corresponding to Fig. 2 is added to the top scale of the figures.}
\end{figure}
The periods of $0.2591$ $day$ and $0.2979$ $day$ correspond approximately to 3.86 and 3.36 cycles per day, respectively. The difference is just 0.5 cycles per day. Lightcurves mainly represent the light scattering cross section of objects. When we assume that an object is a symmetric ellipsoidal body, almost the same cross section appears in every half rotation. Therefore, it is difficult to distinguish the difference of the half rotation using the short observation time, which is comparable with the sidereal rotation period. We call the indistinctive period a pseudo-period. The periods of $0.2591$ $day$ and $0.2294$ $day$ (= 4.36 cycles per day) are the pseudo-period of $0.2979$ $day$. 
Second, the period of $0.2979$ $day$ is able to explain the previous reports about the sidereal rotation period of 107P. Since the period of $0.2979$ day is around twice the period of \citet{Harris83} ($0.1482$ $day$ ${\simeq}$ 3.556 $hour$), their data show enough periodicity in $0.2979$ $day$. Needless to say, assuming the lightcurve of 107P has a triple-peak, the period of $0.1490$ $day$ (${\simeq}$ 3.58 $hour$) is also a candidate for the sidereal rotation period. However, the possibility of $0.1490$ $day$ is eliminated by the inconsistency with the data of \citet{Osip95}. \citet{Harris83} would recognize their lightcurve as the typical double-peak with the period of $3.556$ $hour$, because the third amplitude of flux in their lightcurve was not detected. Moreover, the period of \citet{Osip95} ($0.2542$ $day$ = $6.1$ ${\pm} $ $0.05$ $hour$) is approximately the same as $0.2591$ $day$ (${\simeq}$ $6.22$ $hour$). Our data set also has a sufficiently high significance level around the period of $0.2542$ $day$. As we mentioned above, however, the period of around $0.2591$ $day$ is a pseudo-period. Since the observation term of \Citet{Osip95} was only two days, the demarcation of a pseudo-period would have been difficult. 
Third, we focus on the lightcurve of $0.2979$ $day$ as having an unusual six peaks. The period of $0.0993$ $day$ is just one third that of $0.2979$ $day$. If 107P has a typical double-peak lightcurve, the period of $0.0993$ $day$ is the sidereal rotation period. However, the amplitudes overlap the different three peaks and bottoms in the folded lightcurve with a period of $0.0993$ $day$. Thus, we exclude the period of $0.0993$ $day$ as the sidereal rotation period.

 On the other hand, the period of $0.0993$ $day$ may be the precession period. If an object has tumbling motions, the lightcurve is dominated by two periods: one, $P_{\psi}$, for the rotation about the extremal axis of  the object as an inertia ellipsoid, and the other, $P_{\phi}$, for the precession about the total rotational angular momentum vector. When frequencies are defined as $2f_{\psi}$ = $P^{-1}_{\psi}$ and $2f_{\phi}$ = $P^{-1}_{\phi}$, the lightcurve periodicity of tumbling objects appears at frequencies that are a linear combination of $f_{\psi}$ and $f_{\phi}$ \citep{Kaasa01a}. When we assume that $P_{\psi}$ is $0.2979$ $day$ and $P_{\phi}$ is $0.0993$ $day$, the frequency $4f_{\psi}$ = $2(f_{\phi} - f_{\psi})$ = $6.713$ $day^{-1}$ approximately corresponds to the inverse of period of \cite{Harris83} and one half of our rotational period of $0.2979$ $day$. The existence of periodicity of the linear combination of two periods shows circumstantial evidence for tumbling. We make the shape model of 107P in the  following subsection and discuss the feasibility of tumbling motion in Section 4.1. 
 
\subsection{Direction of total rotational angular momentum and shape model}
Above, we suggested the possibility of tumbling motion. If 107P is a tumbler, the pole orientation does not accord with the direction of total rotational angular momentum and is not stable. What we can obtain is not the pole orientation but the direction of total rotational angular momentum.
 The direction of total rotational angular momentum of 107P can be estimated using the ``epoch method'' \citep{Mag86} or the ``lightcurve inversion method'' (\citealp{Kaasa01b}; \citealp{Kaasa01c, Kaasa02}). The ``amplitude method'' is also proposed as an alternative method (\citealt{Mag86}). However, we cannot adopt the amplitude method because of the small amplitude change during the observational term.
The epoch method determines the direction of total rotational angular momentum by minimizing the phase-shift of the $SF$ in lightcurves. We select a lightcurve peak around the phase of 0.01 at the bottom of Fig. 2 as the $SF$ because the peak is better observed than any other feature. Moreover, we use the data obtained from November 5, 2009 to February 5, 2010. The identification of the lightcurve peak is difficult from the data of another term because of the photometric error. The phase-shift can be written as

\begin{equation}
\frac{T_i-T_0}{P_{\psi}}-n_i=\frac{\theta_i-\theta_0}{2\pi},
\end{equation}
where $T_0$ is the time at the first $SF$, $T_i$ is the time at the $i$-th $SF$, $P_{\psi}$ is the sidereal rotational period, and $n_i$ is the number of rotations between $T_0$ and $T_i$. $\theta_0$ and $\theta_i$ are the projected directions of the phase angle bisector $(PAB)$ in the plane that is perpendicular to the direction of total rotational angular momentum at $T_0$ and $T_i$, respectively. Table 4 shows the epoch of $SFs$. The left hand of Eq. (3) is estimated from the observations; the right hand is theoretically calculated from the tentative direction of total rotational angular momentum and the orbital information of 107P. We define $\delta$ with the following equation  

\begin{equation}
\delta=\sum_{i}^{N}\sqrt{\bigg(\frac{T_i-T_0}{P_{\psi}}-n_i-\frac{\theta_i-\theta_0}{2\pi}\bigg)^2/(N-1)},
\end{equation}
where $N$ is the number of the epoch; here $N$ = $4$. We can estimate the direction of total rotational angular momentum by seeking the minimum of $\delta$. Fig. 4 shows the $\delta$ map that is obtained by scanning the celestial sphere with a trial axis in steps of 1$^{\circ}$ in ecliptic longitude and latitude. Two candidates are found near the directions, A ($\lambda=310^{\circ}$, $\beta=-10^{\circ}$) and  B ($\lambda=132^{\circ}$, $\beta=-17^{\circ}$). Since the epoch method generally derives two solutions with around the same significance level, we cannot determine a unique solution.


\begin{longtable}[c]{cccc} 
\caption{Epoch of Standard Feature $(SF)$ and the amount of phase-shift.}\\
\hline
Year/Mon/Day& Ecliptic longitude  & Ecliptic latitude & Amount of phase-shift\\ 
&(PAB)[${^\circ}$] & (PAB)[${^\circ}$] &  ($\times$10$^{-2}$)  \\
\hline 
2009/11/5   & 341.958 & 4.611 & 0           \\
2009/12/10 & 30.902   & 2.805 & $-5.305$  \\
2009/12/20 & 54.706   & 1.094 & $-6.113$  \\
2010/01/3     &  57.829  & 0.879  & $-7.794$  \\
2010/02/5     &  76.114  & $-0.202$ & $-8.506$ \\

\hline

\end{longtable}

\begin{figure}
\begin{center}
\begin{tabular}{c}
{\includegraphics[width=8cm,clip]{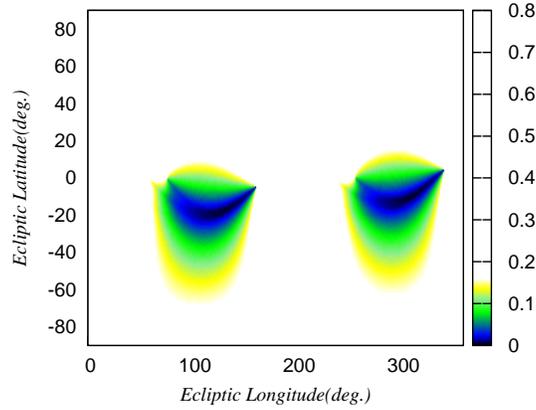}}
\end{tabular}
\end{center}
\caption{$\delta$ maps for the direction of total rotational angular momentum by the epoch method.}
\end{figure}
The lightcurve inversion method derives the most adequate shape model and the corresponding direction of total rotational angular momentum. When carrying out the lightcurve inversion method, we set the initial conditions for the direction of total rotational angular momentum and the sidereal rotation period to that of $0.2979$ $day$. We seek the least deviation between the observational lightcurve and the reconstructed lightcurve from the shape model by scanning the direction of total rotational angular momentum in steps of 1$^{\circ}$ in ecliptic longitude and latitude. We use the data of November 16, 1979 \citep{Harris83}; December 1--2, 1992 \citep{Osip95}; and our high photometric precision data (the error is less than 0.05 mag), which was obtained from November 7, 2009 to January 18, 2010. Fig. 5 shows the deviation map of the direction of total rotational angular momentum using the lightcurve inversion method. 
\begin{figure}
\begin{center}
\begin{tabular}{cc}
{\includegraphics[width=8cm,clip]{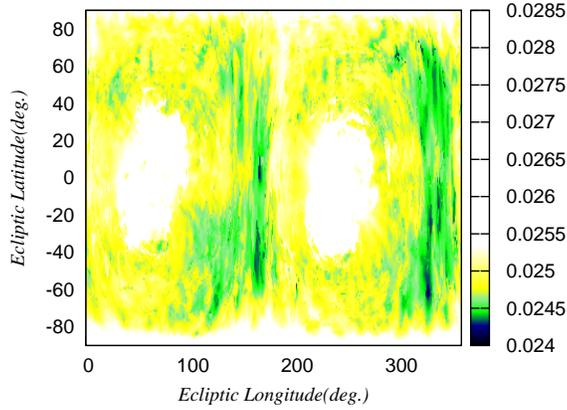}}
\end{tabular}
\end{center}
\caption{Deviation maps for the direction of total rotational angular momentum by the lightcurve inversion method.}
\end{figure}
In addition to the candidates of the epoch method, we have found three other candidates: C ($\lambda=330^{\circ}$, $\beta=-27^{\circ}$), D ($\lambda=328^{\circ}$, $\beta=-61^{\circ}$), and E ($\lambda=167^{\circ}$, $\beta=7^{\circ}$). However, we exclude the candidates D and E because they are less compatible with the epoch method. Note that the observational data was obtained in the phase angle from 21$^{\circ}$ to 77$^{\circ}$. There is no data in a low phase angle. Furthermore, the direction of rotational angular momentum has an uncertainty of typically more than 5$^{\circ}$. For example, though the ground-based observation of Itokawa showed that the pole orientation was $\lambda=355^{\circ}$, $\beta=-84^{\circ}$ \citep{Kaasa03},  the Hayabusa spacecraft revealed that the pole orientation of Itokawa was $\lambda=128.5^{\circ}$, $\beta=-89.66^{\circ}$ \citep{Demura06}. Moreover, we can see from Table 4 that the $SF$ in the lightcurve shifts to the negative direction with time. As a corollary, the direction of total rotational angular momentum of the three candidates is south of the ecliptic plane. This indicates that the sense of sidereal rotation is retrograde. The lightcurve of Fig. 6 has been calibrated for the phase-shift. The six peaks and the periodicity in the lightcurve become clearer than in the lightcurve of $0.2979$ $day$ period in  Fig. 2. This result adds to the evidence of the retrograde rotation.

\begin{figure}
\begin{center}
\begin{tabular}{c}
{\includegraphics[width=8cm,clip]{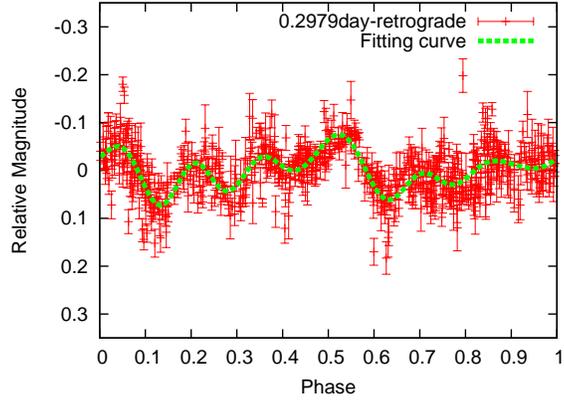}}
\end{tabular}
\end{center}
\caption{Calibrated lightcurve for phase-shift. The fitting curve is described by two-order two-dimensional Fourier series. }

\end{figure}
 Next, we make the three shape models of 107P for the directions of total rotational angular momentum A, B and C. The shape model A is shown in Fig. 7.  
\begin{figure}[htbp]
  \begin{center}
    \begin{tabular}{c}
      \begin{minipage}{0.33\hsize}
        \begin{center}
          \includegraphics[clip, width=4.45cm]{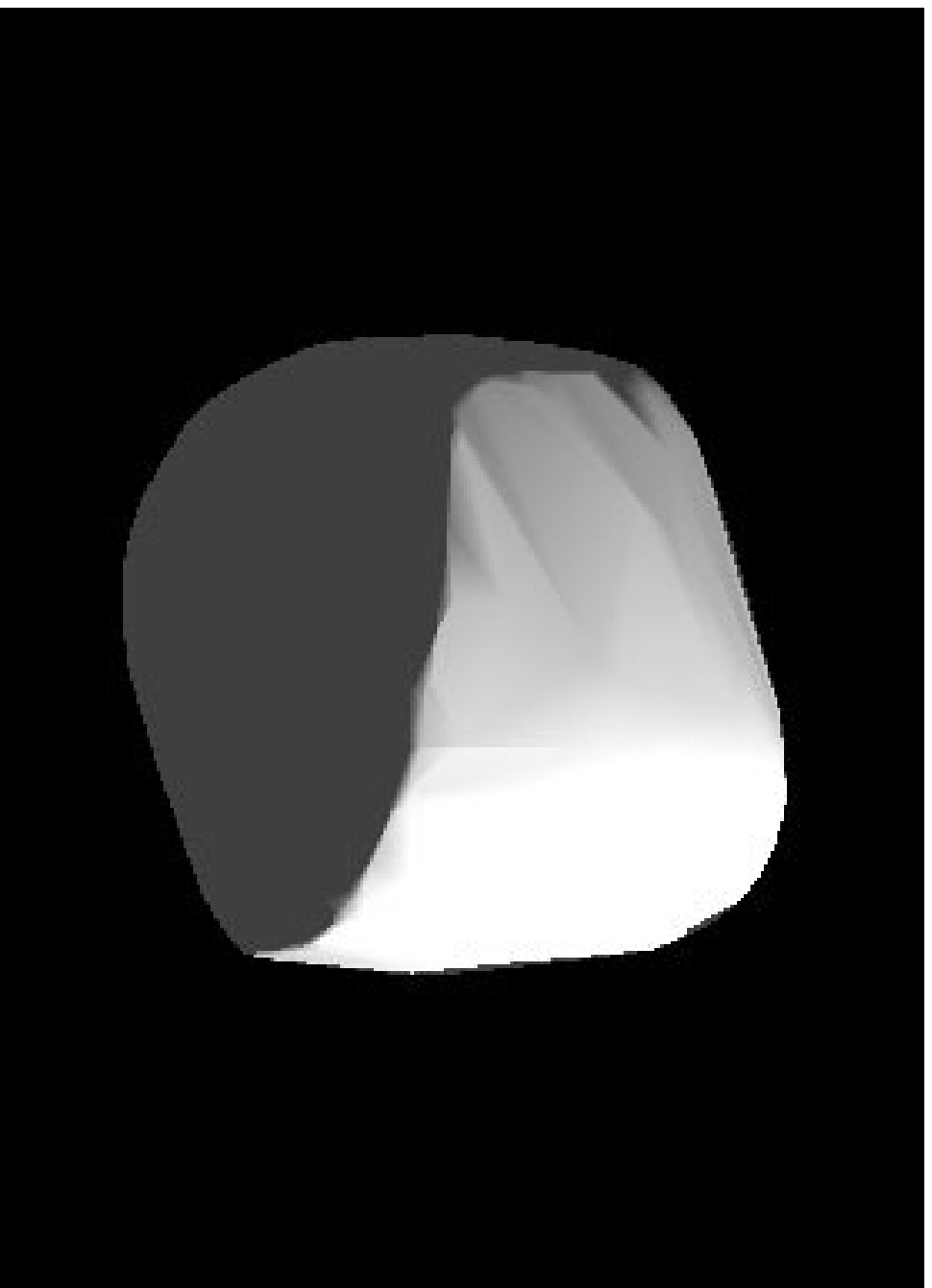}
        \end{center}
      \end{minipage}
      \begin{minipage}{0.33\hsize}
        \begin{center}
          \includegraphics[clip, width=4.48cm]{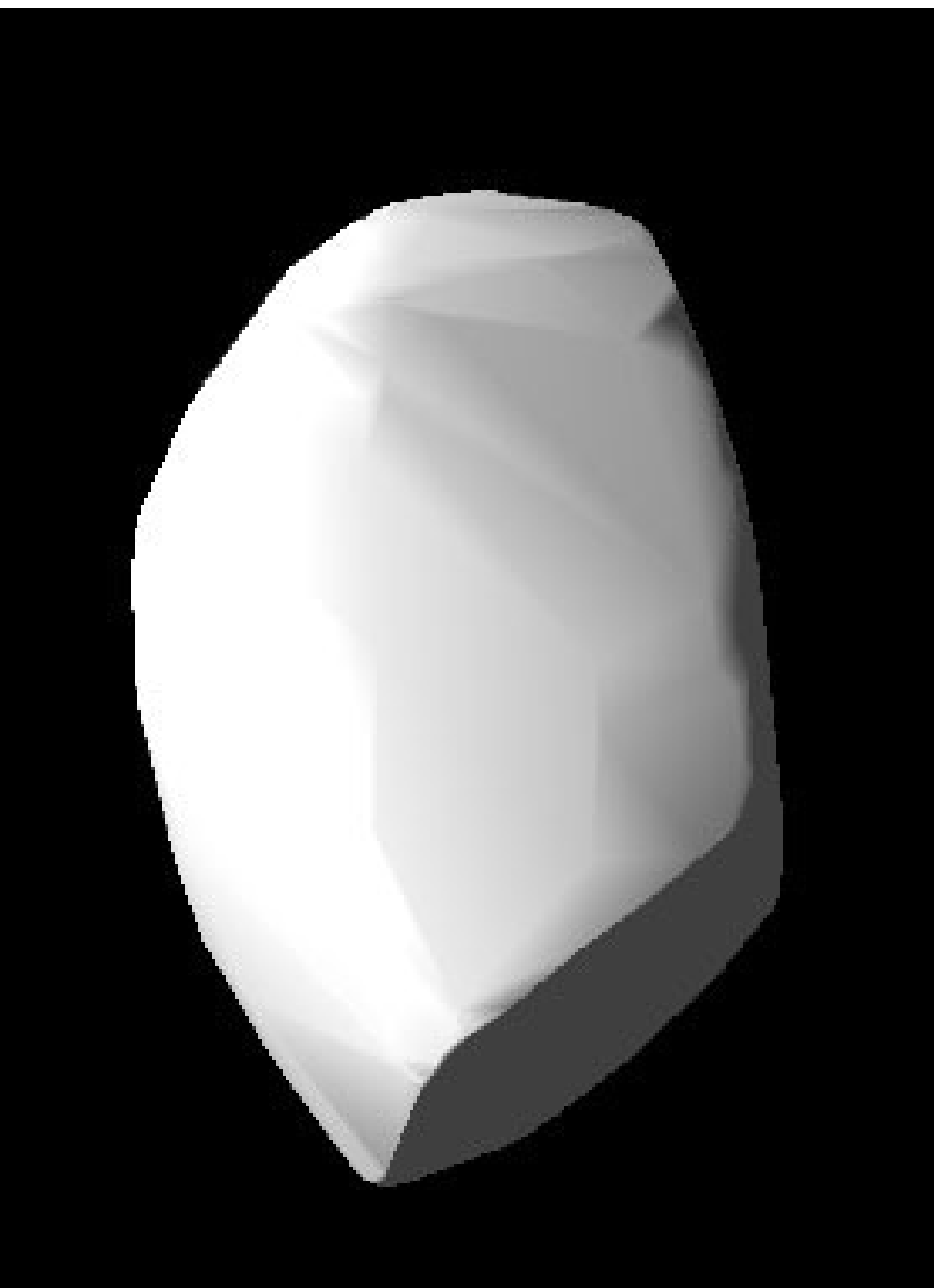}
        \end{center}
      \end{minipage}
      \begin{minipage}{0.33\hsize}
        \begin{center}
          \includegraphics[clip, width=4.5cm]{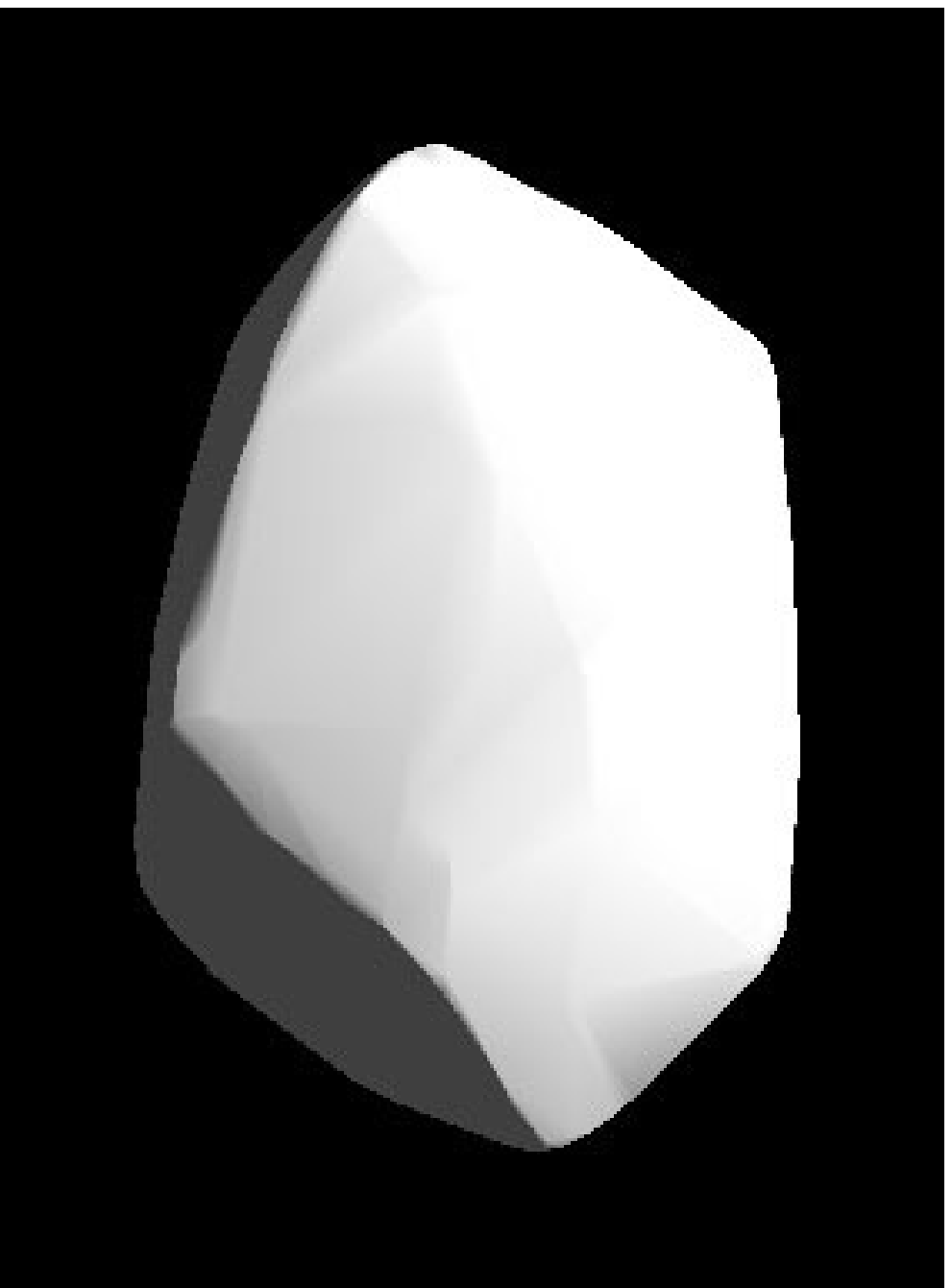}
        \end{center}
      \end{minipage}
    \end{tabular}
    \caption{Shape model A of 107P. (Left) Pole-on view. (Center) Equatorial view from the right side of the pole-on image. (Right) Equatorial view from the bottom of the pole-on image. }
  \end{center}
\end{figure}
Here, $L_{1}$, $L_{2}$, and $L_{3}$ are defined as the normalized axis length when 107P is a triaxial ellipsoid body. The axes satisfy the relationship $L_{1}$ ${\leq}$ $L_{2}$ ${\leq}$ $L_{3}$. $L_{3}$ is a sidereal rotation axis of the shape model A. We have found that the normalized axis lengths $L_{1}$, $L_{2}$, and $L_{3}$ are around 1.0, 1.0, and 1.6, respectively. The axis ratio of the shape model B is around the same as that of the shape model A. The shape model A and B indicate a so-called long axis mode (LAM).  Some previous studies have described the motion of a force-free asymmetric rigid body (\citealp{Samarasinha91}; \citealp{Kaasa01a}). Now, $L_{1}$ $\simeq$ $L_{2}$ indicates that the equations of force-free precession are simplified to the following, 
 
 \begin{equation}
\dot{\psi}=\cos\theta\left(\frac{M}{I_{3}}-\dot{\phi}\right),
\end{equation}
 
\begin{equation}
\dot{\phi}=\frac{M}{I_{1}},
\end{equation}

\begin{equation}
I_{1}=\frac{\mu}{20}(L_{2}^{2}+L_{3}^{2}),
\end{equation}

\begin{equation}
I_{3}=\frac{\mu}{20}(L_{1}^{2}+L_{2}^{2}).
\end{equation}
Here,  $\psi$, $\phi$, $\theta$ are the Euler angles of sidereal rotation, precession, and nutation, respectively. $M$ is the total rotational angular momentum in an inertial frame. $I_{1}$ and $I_{3}$ express the inertia moment of a triaxial ellipsoid by using mass $\mu$. Moreover, the equations show that the motion of the external axis about $M$ occurs as a constant rate. From these equations and $\dot{\phi}$ = 3$\dot{\psi}$, the nutation is negligible, and the angle $\theta$ is constant around $65^{\circ}$. A tilted, rugby-ball-shaped body rotates with a period of $0.0993$ $day$ about the total rotational angular momentum, and with a period of $0.2979$ $day$ about the external axes of 107P itself. Alternatively, we can assume a case that has the sidereal rotation of $0.0993$ $day$ and the precession period of $0.2979$ $day$. Substituting $\dot{\psi}$ = 3$\dot{\phi}$, there is no solution for the nutation angle. Therefore, the assumption is not adequate.

 Meanwhile, as we show in Fig. 8, the normalized axis lengths for the shape model C are around 1.5, 1.5, and 1.0.  Here, the axes satisfy the relationship $L_{1}$ ${\geq}$ $L_{2}$ ${\geq}$ $L_{3}$. $L_{3}$ is a sidereal rotation axis. Thus, the shape model C is a short axis mode (SAM). The rotation and precession of a SAM are in opposite direction from $I_{1}$ $<$ $I_{3}$. We calculate the nutation angle by assuming $\dot{\phi}$ = $-3$$\dot{\psi}$. However, there is no solution for the nutation angle. To obtain the solution for the nutation angle, $\dot{\phi}$ should be less than $-3.62$$\dot{\psi}$ if the axis lengths of the shape model C are correct, or the axis lengths of $L_{1}$ and $L_{2}$ should be longer than  $\sqrt{\mathstrut 3}$ $L_{3}$ if $\dot{\phi}$ = $-3$$\dot{\psi}$ and $L_{1}$ $\simeq$ $L_{2}$ are correct. These situations are inconsistent with our results. Therefore, 107P of the shape model C is a non-precession object rather than a precessional object. 

 \begin{figure}[htbp]
  \begin{center}
    \begin{tabular}{c}
      \begin{minipage}{0.33\hsize}
        \begin{center}
          \includegraphics[clip, width=4.5cm]{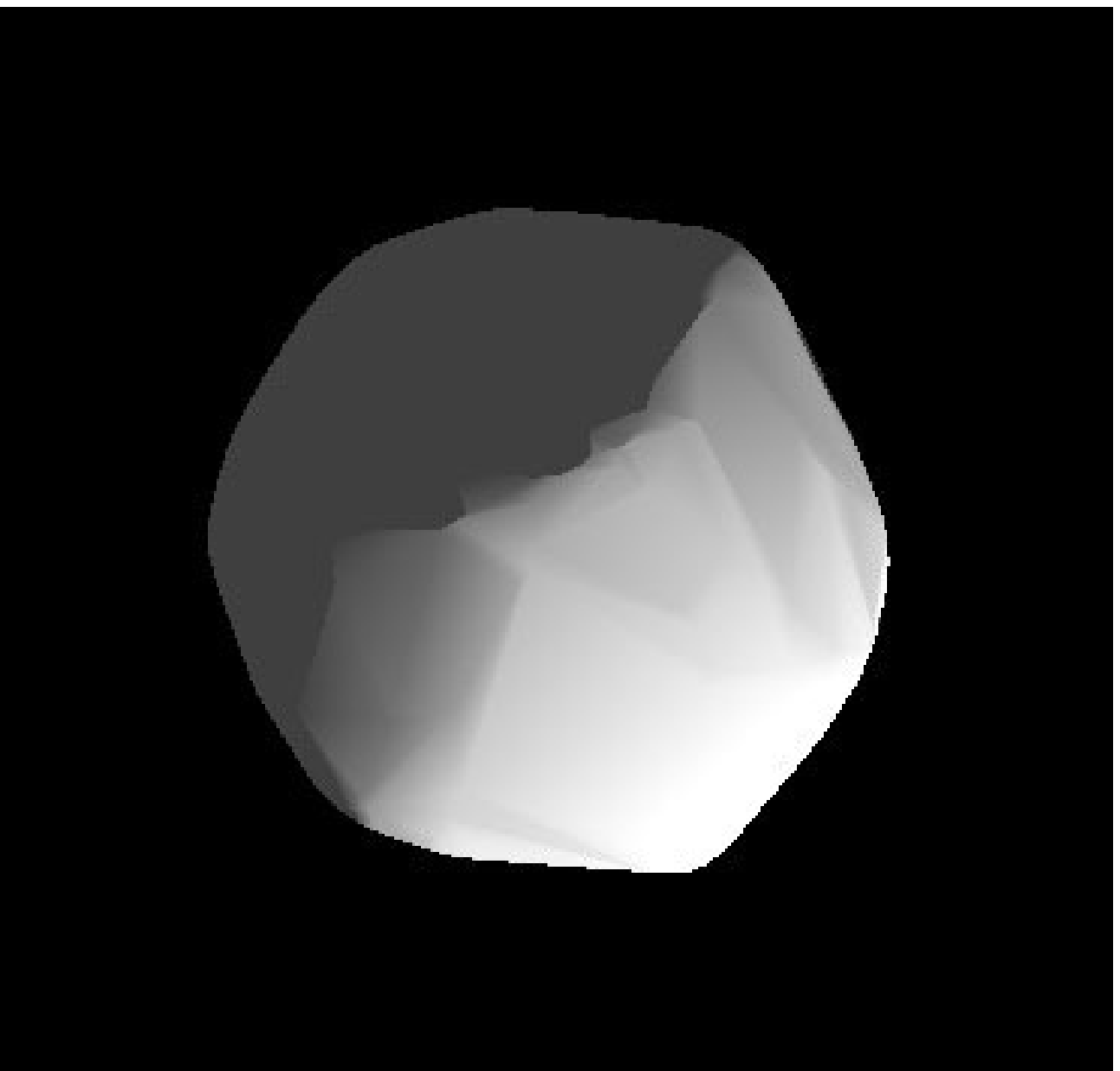}
        \end{center}
      \end{minipage}
      \begin{minipage}{0.33\hsize}
        \begin{center}
          \includegraphics[clip, width=4.5cm]{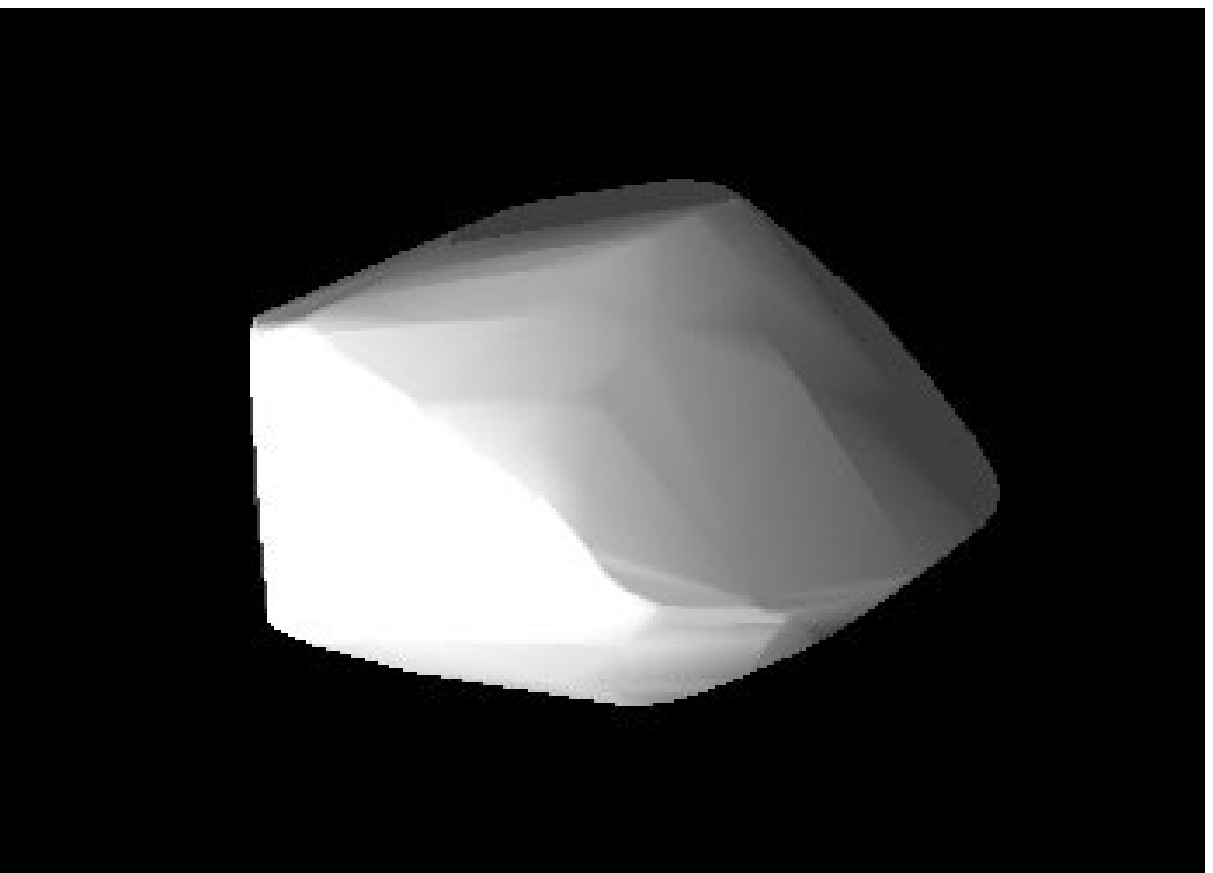}\\
        \end{center}
      \end{minipage}
      \begin{minipage}{0.33\hsize}
        \begin{center}
          \includegraphics[clip, width=4.5cm]{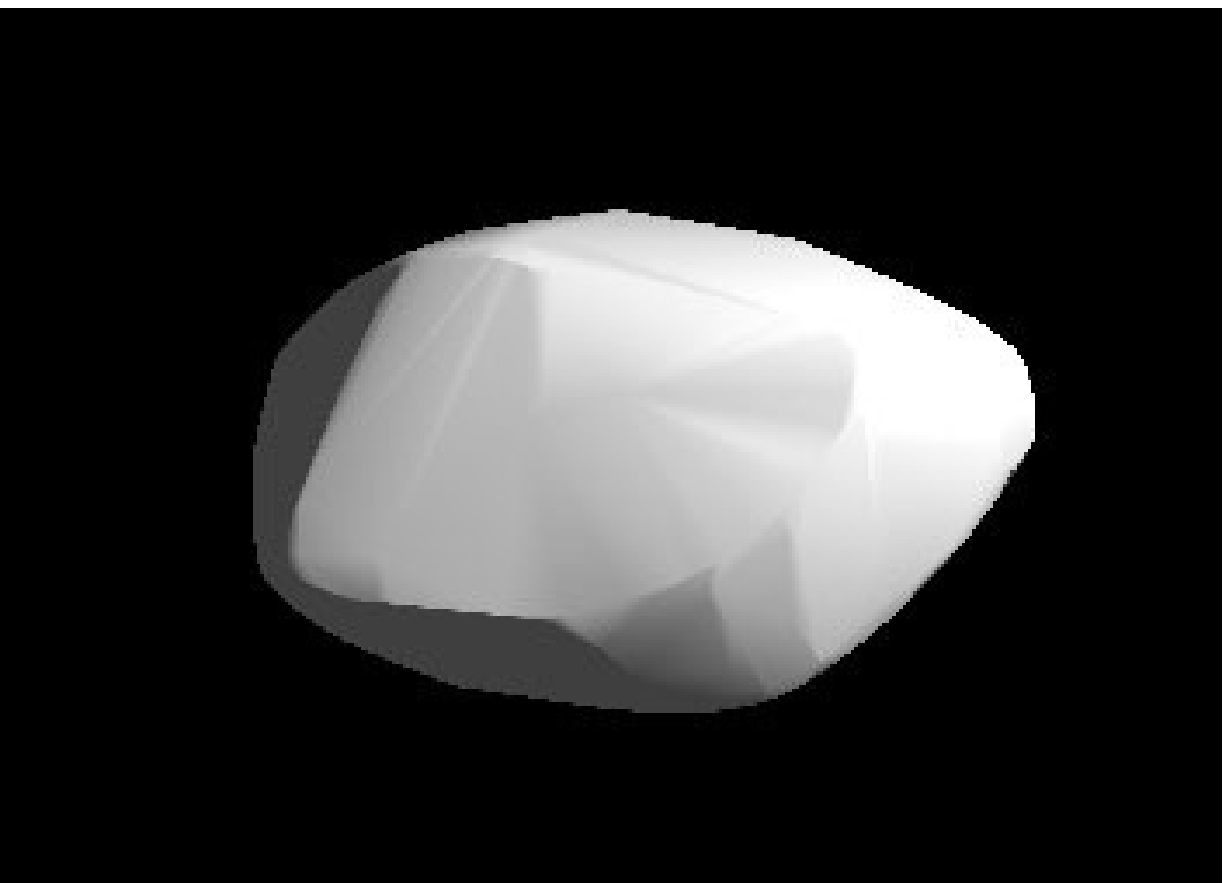}
        \end{center}
      \end{minipage}
    \end{tabular}
    \caption{Shape model C of 107P. (Left) Pole-on view. (Center) Equatorial view from the right side of the pole-on image. (Right) Equatorial view from the bottom of the pole-on image. }
  \end{center}
\end{figure}
 
\subsection{Taxonomic class and rotational color variations}
The taxonomic class and rotational color variations for 107P were investigated by a color--color diagram. We note that  the classification of subclasses, such as B, F, or G-type, is difficult using multiband photometry. We conducted the multiband photometry eight times ($Phase$-$1$ to -$8$). The obtained color--color diagram and the color index are shown in Fig. 9 and Table 5. We utilized the $z'$ images of $Phase$-$6$ for those of $Phase$-$7$ due to the poor weather conditions in $Phase$-$7$. The color--color diagram macroscopically shows that 107P is a C-type (including B, F, G-type) asteroid. The colors of $Phase$-$2$ and $Phase$-$6$ indicate typical C-type features in the three color indices. The others are slightly reddish features like an X-type asteroid in the color index of $r'-i'$. Though only the $g'-r'$ of $Phase$-$3$ barely exceed the one-sigma of mean color index in Table 5, it is difficult to assert the detection of the rotational color variation due to the photometric error. In addition to it, the long observation term of $\sim$0.15 in phase ($\sim$1.0 hour) for each sequence obscures the detection of rotational color variation.
In order to confirm the color variations, follow-up observations and/or exploration by spacecraft are needed. 
\newpage
\begin{scriptsize}
\begin{longtable}[c]{cccccc} 
\caption{Color index of 107P.  The observation term (Obs term) of each sequence is expressed as the rotational phase in the lightcurve. Since the data of the OAO are obtained with three bands, there is no color index of $i'-z'$. Mean shows the arithmetic average and standard deviation of each color index.}
\\
\hline
 &  Obs term [Phase] & Observatory & $g' - r'$ & $r' - i'$  & $i' - z'$   \\ 

\hline 
Phase-1& 0.0365--0.1221 & OAO& 0.409 ${\pm}$ 0.036 & 0.217 ${\pm}$ 0.036  &  \\ 
Phase-2& 0.0048--0.1590 & BSGC& 0.462 ${\pm}$ 0.055 & 0.141 ${\pm}$ 0.043  & 0.039 ${\pm}$ 0.044  \\
Phase-3& 0.1348--0.2177 & OAO& 0.522 ${\pm}$ 0.037 & 0.184 ${\pm}$ 0.036  &  \\
Phase-4& 0.1941--0.3335 & BSGC& 0.451 ${\pm}$ 0.056 & 0.190 ${\pm}$ 0.034  & 0.018 ${\pm}$ 0.049  \\
Phase-5& 0.2345--0.3177 & OAO& 0.364 ${\pm}$ 0.040 & 0.177 ${\pm}$ 0.039  &  \\
Phase-6& 0.3461--0.5045 & BSGC& 0.444 ${\pm}$ 0.057 & 0.138 ${\pm}$ 0.045  &0.050 ${\pm}$ 0.043    \\
Phase-7& 0.4662--0.5426 & BSGC& 0.382 ${\pm}$ 0.100 & 0.173 ${\pm}$ 0.054  & 0.021 ${\pm}$ 0.053 \\
Phase-8& 0.9590--1.0239 & OAO& 0.423 ${\pm}$ 0.040 & 0.176 ${\pm}$ 0.040  &  \\
\hline
Mean  & --- & --- & 0.432  ${\pm}$ 0.050 & 0.175 ${\pm}$ 0.026  &  0.032 ${\pm}$ 0.015    \\
\hline
\end{longtable}
\end{scriptsize}

\begin{figure}
\begin{center}
\begin{tabular}{c}
{\includegraphics[width=8cm,clip]{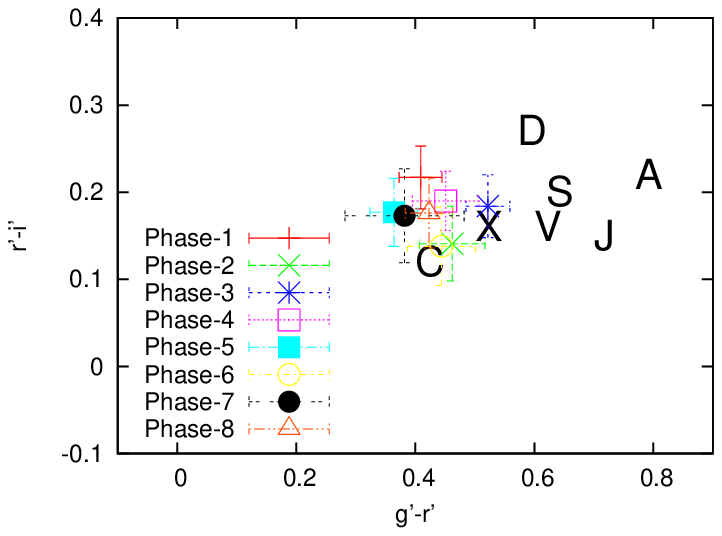}}\\
{\includegraphics[width=8cm,clip]{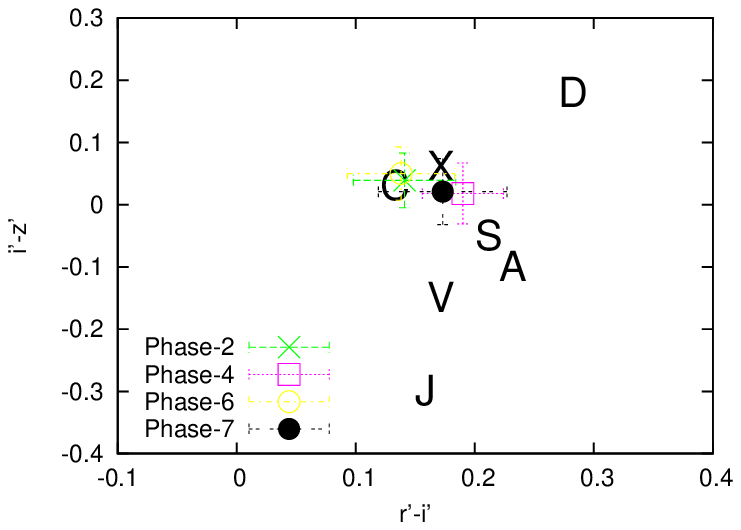}}
\end{tabular}
\end{center}
\caption{Color--color diagram of 107P. Letters in the figure represent the taxonomic classes of asteroids on the color--color diagram \citep{Ive01}. X-type asteroids include E, M, and P-type asteroids. The case of low-albedo asteroids indicates P-type.  }
\end{figure}

\newpage

\section{Discussion}
\subsection{Tumbling}
We discuss the possibility of tumbling. There were some reports in which asteroid lightcurves  indicated tumbling, e.g.,  (253) Mathilde \citep{Mottola95}, (3288) Seleucus \citep{Harris99}, and (4179) Toutatis (\citealp{Spencer95}; \citealp{Kry99}). \cite{Pravec05} assessed the validity of tumbling for these asteroids based on whether the lightcurves could be approximated with two-dimensional Fourier series and the physical model of tumbling could be constructed. The two-dimensional Fourier series is described in the following form 

\begin{eqnarray}
\nonumber  F^{m}(t)                 & =& C_{0}+\sum^{m}_{j=1}\left[C_{j0}\cos\frac{2{\pi}j}{P_{\psi}}t+S_{j0}\sin\frac{2{\pi}j}{P_{\psi}}t\right]\\
\nonumber            &+& \sum^{m}_{k=1}\sum^{m}_{j=-m}\left[C_{jk}\cos\left(\frac{2{\pi}j}{P_{\psi}}+\frac{2{\pi}k}{P_{\phi}}\right)t\right.\\
            &+& \left. S_{jk}\sin\left(\frac{2{\pi}j}{P_{\psi}}+\frac{2{\pi}k}{P_{\phi}}\right)t\right],
\end{eqnarray}
where $m$ is the order, $C_{0}$ is the mean reduced light flux, $C_{jk}$ and $S_{jk}$ are the Fourier coefficients for the linear combination of the two frequencies $P^{-1}_{\psi}$ and $P^{-1}_{\phi}$, respectively, and $t$ is the time. Substituting $m$ = $2$, $P_{\psi}$ = $0.2979$ $day$, and $P_{\phi}$ = $0.0993$ $day$ for 107P, we obtain a fitting curve, as shown in Fig. 6. The fitting curve adequately reconstructs the trend of the lightcurve.  However, since the sidereal rotation period and the processing period have a commensurability of 3:1, the lightcurve can also be reconstructed using the one-dimensional Fourier series of sixth order. As we mentioned in Section 3.2, the physical model is possibly constructed using a LAM of $L_{1}$:$L_{2}$:$L_{3}$ = 1.0:1.0:1.6, ($\lambda=310^{\circ}$, $\beta=-10^{\circ}$) or ($\lambda=132^{\circ}$, $\beta=-17^{\circ}$), $\theta$ = $65^{\circ}$, $P_{\psi}$ = $0.2979$ $day$, and $P_{\phi}$ = $0.0993$ $day$. Although \cite{Pravec05} mentions that a tumbling asteroid generally does not return to the same orientation in any single period, the approximately equal length of $L_{1}$ and $L_{2}$ for 107P suggests a negligible change for the nutation angle. Therefore, 107P can return to the same orientation every $0.2979$ $day$. These circumstances imply that 107P might be a tumbling object. 

Assuming 107P is a tumbler, external forces are required to trigger the motion. Impacts of small objects, tidal encounters with planets, and YORP effects are suggested by \cite{Pravec05}. Though 107P is a NEO, the object did not have an encounter with Earth in 1949. In the case of km-size objects, the efficient onset of tumbling by YORP requires a longer timescale than that of collision with small objects \citep{Vok07}. Therefore, we propose the impact of small objects as a probable cause for tumbling of 107P. The orbital origin of 107P has a high possibility of being from the outer MBA region inhabited by MBCs. One possibility is that the cometary activities of MBCs  are caused by impacts of small objects. We can consider that 107P is originally an object like an MBC and impacts with small objects in the NEO region could eject dust and/or expose sub-surface ice that then trigger 107P's cometary activity. When we suppose that the collisional excitation happened in 1949, the damping timescale \citep{Harris94} is expressed as         

\begin{equation}
\tau=\frac{P^{3}_{\psi}}{C^{3}D{^2}},
\end{equation}
where $D$ is the mean diameter of tumblers in kilometer units and $C$ is a constant of about 17 (uncertain by about a factor of 2.5). The units of $P_{\psi}$ and $\tau$ are hours and billion ($10^{9}$) of years, respectively. Since the damping timescale of around $6.2 \times 10^{6}$ $yr$ is long enough, 107P would continue tumbling even if the impact occurred before 1949. 

\subsection{Binary asteroids}
We describe the situation in which 107P hosts a binary. In order to confirm the existence of a binary, the detection of mutual eclipse events is required in the lightcurve. The mutual eclipse events were not detected in the observations of \cite{Harris83} and \cite{Osip95} because of the viewing angle, the lower photometric precision, or the absence of the binary.  On the other hand, we detected the around same flux decrease in every 0.50 phase. Therefore, the existence of the binary is conceivable as the other interpretation of the shape model C. If we define the flux decrease around the phase of 0.15 (or 0.30, 0.45) and 0.65 (or 0.80, 0.95) in Fig. 6 as the primary (secondary) eclipse and the secondary (primary) eclipse, respectively, the orbital period of the binary is $0.2979$ $day$. Supposing a circular orbit and negligible mass for the binary, the semi-major axis is described as

\begin{equation}
a=\left(\frac{GMP^{2}_{orb}}{4{\pi}^{2}}\right)^{\frac{1}{3}},
\end{equation}
where $G$ is the gravitational constant, $M$ is the mass of 107P, and $P_{orb}$ is the orbital period of the binary. For the sake of simplicity, when assuming that 107P is spherical with the diameter of 3.46 $km$ \citep{Licandro09} and a typical density of 2 $g/cm^{3}$, the semi-major axis is around 3.65 $km$. In the case of the same albedo for 107P and the binary, the flux decrease of the total eclipse ($A_{mut}$)  satisfies the following relationship \citep{Polishook11} 

\begin{equation}
A_{mut}=2.5 log\left[1+\left(\frac{R_{s}}{R_{p}}\right)^{2}\right],
\end{equation}
where $R_{s}$ is the radius of the binary and $R_{p}$ is the radius of 107P. Since the typical flux decrease is $\sim$0.05 $mag$ in Fig. 6, the radius of  the binary is around 0.4 $km$. When we assume the orbital plane of 107P accords with the line of sight from an observer, the inclination of the binary as an occulter satisfies

\begin{equation}
\sin i < \frac{R_{p}+R_{s}}{a}.
\end{equation}
Here, $i$ is the inclination of the binary for the orbital plane of 107P is less than 36$^{\circ}$ in the 107P system. If $i$ is zero, the eclipse duration is estimated to be $\sim$0.05 $day$. The term is around one-sixth of the orbital period and consistent with the interval of lightcurve peaks of Fig. 6. Moreover, the binary hypothesis indicates that the double-peak period of the lightcurve without the eclipse becomes $0.1490$ $day$. As we mentioned in Section 3.1, the period of $0.1490$ $day$ as the sidereal rotation of 107P is not likely. Alternatively, the lightcurve without the eclipse might be a quadruple-peak lightcurve whose period is $0.2979$ $day$. Though the quadruple-peak lightcurve is rare, the period could compatibly account for all the past reports. In addition, the situation shows that the sidereal rotation of 107P and the orbital periods of the binary are locked with $0.2979$ $day$. The period of $0.0993$ $day$ is explained by the period between the egress time of the primary (secondary) eclipse and the ingress time of the secondary (primary) eclipse. 

The promising mechanisms for formation of asteroid binaries are the rotational-fission due to the spin-up by YORP effects (\citealp{Scheeres07}; \citealp{Pravec07}; \citealp{Walsh08}), tidal encounter with planets (\citealp{Richardson98}; \citealp{Bottke99}; \citealp{Walsh06}), and the escaping ejecta by the collisions (\citealp{Durda04}; \citealp{Polishook11}). The mechanisms have a lot in common with the cause of  tumbling. Fissions and collisions in every mechanism can trigger 107P's cometary activity. The possible existence of a binary is consistent with the past cometary activity.

\section{Summary}
This study revealed the physical properties of 107P by a photometric observation campaign. We detected the lightcurve periodicity to be $0.2979$ $day$ and $0.0993$ $day$ with a commensurability of 3:1. The multiband photometry indicates that the taxonomy class of 107P is C-type. No clear rotational color variations are confirmed on the surface. We suggested two models to explain the different interpretations of the lightcurve periodicity.

1. The commensurability reflects tumbling with the sidereal rotation period of $0.2979$ $day$ and the precession period of $0.0993$ $day$. The shape is a LAM of $L_{1}$:$L_{2}$:$L_{3}$ = 1.0:1.0:1.6. Around the same length of $L_{1}$ and $L_{2}$ shows the nutation angle is approximately constant at $65^{\circ}$. The direction of total rotational angular momentum is around $\lambda=310^{\circ}$, $\beta=-10^{\circ}$, or $\lambda=132^{\circ}$, $\beta=-17^{\circ}$. 107P returns to the same orientation every $0.2979$ $day$ by retrograde motion. Impacts of small objects are suggested as a cause for the tumbling and comet activity. Alternatively, the past comet activity itself is thought to be a cause of the tumbling, like a 1P/Halley \citep{Samarasinha91}.

2. 107P is not a tumbler. The sidereal rotation period is $0.2979$ $day$. The shape is roughly spherical but slightly hexagonal with a SAM of $L_{1}$:$L_{2}$:$L_{3}$ = 1.5:1.0:1.0. The pole orientation is around $\lambda=330^{\circ}$, $\beta=-27^{\circ}$. The sense of rotation is retrograde. The lightcurve of commensurability  would reflect a discriminative appearance like (2867) Steins, which has been explored by the Rosetta spacecraft \citep{Keller10}. Otherwise, the lightcurve also indicates the possibility of hosting a binary whose orbital period is $0.2979$ $day$. The existence of a binary is also consistent with the past cometary activity.

Finally, we describe the mission feasibility for 107P.  The orbit accessible by spacecraft makes 107P a promising  target for a sample-return mission. If 107P is not a tumbling object, the moderate rotational period of $0.2979$ $day$ would enable us to obtain a sample by the touchdown of a spacecraft, whereas touchdown on 107P would require a difficult maneuver if 107P is a tumbler. In that case, a multi-fly-by mission that combines with the sample return mission for another target would become a hopeful plan.

\section*{Acknowledgments}
We appreciate the observational campaign  participants, especially Dr. Ryosuke Nakamura and Dr. Masanao Abe, for their  dedicated coordination of the campaign and encouragement. We also acknowledge the Japan Space Forum. This work was financially supported in part by a Grant-in-Aid for Scientific Research (KAKENHI)  from the Ministry of Education, Culture, Sports, Science and Technology of Japan (Nos. 22916007, 14GS0211, and 1947003). Sunao Hasegawa was supported by the Space Plasma Laboratory, ISAS, JAXA.

Funding for the SDSS and SDSS-II has been provided by the Alfred P. Sloan Foundation, the Participating Institutions, the National Sci- ence Foundation, the U.S. Department of Energy, the National Aeronautics and Space Administra- tion, the Japanese Monbukagakusho, the Max Planck Society, and the Higher Education Fund- ing Council for England. The SDSS Web Site is http://www.sdss.org/.

The SDSS is managed by the Astrophysical Research Consortium for the Participating Insti- tutions. The Participating Institutions are the American Museum of Natural History, Astro- physical Institute Potsdam, University of Basel, University of Cambridge, Case Western Reserve University, University of Chicago, Drexel Univer- sity, Fermilab, the Institute for Advanced Study, the Japan Participation Group, Johns Hopkins University, the Joint Institute for Nuclear As- trophysics, the Kavli Institute for Particle As- trophysics and Cosmology, the Korean Scien- tist Group, the Chinese Academy of Sciences (LAMOST), Los Alamos National Laboratory, the Max-Planck-Institute for Astronomy (MPIA), the Max-Planck-Institute for Astrophysics (MPA), New Mexico State University, Ohio State Uni- versity, University of Pittsburgh, University of Portsmouth, Princeton University, the United States Naval Observatory, and the University of Washington.

\end{document}